\documentclass[11pt,a4paper]{article}
\usepackage{jcappub}

\usepackage[latin1]{inputenc}
\usepackage{graphicx}
\usepackage{epsfig}
\usepackage{color}
\usepackage{comment}
\usepackage{slashed}

\title{Constraining the Evolution of the Baryon Fraction in the IGM with FRB and $H(z)$ data}

\author[a,b]{Jun-Jie Wei,}
\author[c]{Zhengxiang Li,}
\author[c]{He Gao,}
\author[a,d]{Xue-Feng Wu}

\affiliation[a]{Purple Mountain Observatory, Chinese Academy of Sciences, Nanjing 210034, China}
\affiliation[b]{Guangxi Key Laboratory for Relativistic Astrophysics, Guangxi University, Nanning 530004, China}
\affiliation[c]{Department of Astronomy, Beijing Normal University, Beijing 100875, China}
\affiliation[d]{School of Astronomy and Space Sciences, University of Science and Technology of China, Hefei 230026, China}

\emailAdd{jjwei@pmo.ac.cn; zxli918@bnu.edu.cn; gaohe@bnu.edu.cn; xfwu@pmo.ac.cn}

\abstract{Extragalactic dispersion measures (DMs) obtained from observations of fast radio bursts (FRBs) are an excellent tool
for probing intergalactic medium (IGM) and for conducting cosmography. However, the DM contribution from the IGM
(${\rm DM_{IGM}}$) depends on the fraction of baryon mass in the IGM, $f_{\rm IGM}$, which is not properly constrained.
As $f_{\rm IGM}(z)$ is geometrically related to the Hubble parameter $H(z)$ and ${\rm DM_{IGM}}(z)$, here we propose
that combining two independent measurements of FRBs and $H(z)$ in similar redshift ranges provides a novel and cosmology-free
method to constrain the evolution of $f_{\rm IGM}(z)$. Under the assumption that $f_{\rm IGM}$ is evolving with redshift
in a functional form, we forecast that the evolution of $f_{\rm IGM}(z)$ can be well inferred in a combined analysis of
$\sim3000$ ${\rm DM_{IGM}}(z)$ derived from FRBs and $\sim50$ $H(z)$ derived from the Hubble parameter data.
Though the efficiency of our method is not as good as that of the other model-independent method involving the joint measurements
of DM and luminosity distance of FRBs, our method offers a new model-independent way to constrain $f_{\rm IGM}(z)$.}


\arxivnumber{1907.09772}

\begin{document}
\maketitle

 \flushbottom
%%%%%%%%%%%%%%%%%%%%%%%%%%%%%%%%%%%%%%%%%%%%%%%%%%%%%%%%%%%%%%%%%%%%%%%%%%%%%%%%
%%%%%%%%%%%%%%%%%%%%%%%%%%%%%%%%%%%%%%%%%%%%%%%%%%%%%%%%%%%%%%%%%%%%%%%%%%%%%%%%

%%%%%%%%%%%%%%%%%%%%%%%%%%%%%%%%%%%%%%%%%%%%%%%%%%%%%%%%%%%%%%%%%%%%%%%%%%%%%%%%
%%%%%
%%%%%%%%%%%%%%%%%%%%%%%%%%%%%%%%%%%%%%%%%%%%%%%%%%%%%%%%%%%%%%%%%%%%%%%%%%%%%%%%
%%%%%

\section{Introduction}

Fast radio bursts (FRBs) are a mysterious class of millisecond-duration radio transients, all with
dispersion measures (DMs) in excess of Galactic expectations, signaling an extragalactic or even a
cosmological origin \citep{2007Sci...318..777L,2013Sci...341...53T,2015MNRAS.447..246P,2016PASA...33...45P}.
The localization of the repeating event FRB 121102 at $z=0.19$ confirmed the cosmological origin of at least this source
\citep{2016ApJ...833..177S,2016Natur.531..202S,2017Natur.541...58C,2017ApJ...834L...8M,2017ApJ...834L...7T}.
Very recently, the host galaxies and distances of non-repeating FRBs have been identified as well.
Ref. \cite{2019arXiv190611476B} reported the interferometric localization of the non-repeating burst FRB 180924
to a position 4 kpc from the center of an early-type spiral galaxy at a cosmological redshift of 0.3214.
Similarly, ref. \cite{2019arXiv190701542R} reported the localization of FRB 190523 to a few-arcsecond region
containing a massive galaxy at redshift 0.66.
If lots of FRBs with known redshifts are detected, the combined DM and $z$ information of these events
can be used to measure the baryon number density of the universe \citep{2014ApJ...783L..35D,2016Natur.530..453K},
find circumgalactic baryons \citep{2014ApJ...780L..33M,2017ApJ...834...13F,2018PhRvD..98j3518M,2019ApJ...872...88R},
probe the reionization history of the universe \citep{2014ApJ...783L..35D,2014ApJ...797...71Z},
constrain cosmological parameters \citep{2014ApJ...788..189G,2014PhRvD..89j7303Z,2016ApJ...830L..31Y,2018ApJ...856...65W,2018ApJ...860L...7W},
test the weak equivalence principle \citep{2015PhRvL.115z1101W,2016ApJ...820L..31T},
constrain the rest mass of the photon \citep{2016ApJ...822L..15W,2016PhLB..757..548B,2017PhLB..768..326B,2017PhRvD..95l3010S},
determine the cosmic proper distance \citep{2017A&A...606A...3Y}, and probe compact dark matter \citep{2016PhRvL.117i1301M,2018A&A...614A..50W}
or measure Hubble constant and cosmic curvature \citep{2018NatCo...9.3833L} through gravitational lensing.

However, one issue that stand out for restricting the cosmological applications of FRBs is the strong degeneracy
between cosmological parameters and the fraction of baryon mass in the intergalactic medium (IGM), $f_{\rm IGM}$.
Moreover, $f_{\rm IGM}$ is a poorly known parameter.
The baryon distribution of the universe has been an important subject of study for a long time.
Ref. \cite{1998ApJ...503..518F} presented an estimate of the global budget of baryons in all states based on all
relevant information they had been able to marshal. They showed that stars and their remnants comprise only
about 17\% of the baryons. Although many other researches on the baryon distribution through numerical simulations
\cite{1999ApJ...514....1C,2006ApJ...650..560C,2009RvMP...81.1405M} or observations
\cite{2004ApJ...616..643F,2012ApJ...759...23S,2016PhRvL.117e1301H,2018PhRvD..98j3518M}
have been performed, the baryon fraction in the IGM, $f_{\rm IGM}$, is still not well constrained.
Numerical simulations suggested that at $z\geq1.5$, $\sim90\%$ of the baryons produced by
the Big Bang are contained within the IGM (i.e., $f_{\rm IGM}\approx 0.9$), with only $\sim10\%$ in galaxies, galaxy clusters
or possibly locked up in compact stars \citep{2009RvMP...81.1405M}. While $18\pm4\%$ of the baryons are found to
exist in galaxies, circumgalactic medium, intercluster medium, and cold neutral gas at redshifts $z\leq0.4$ \citep{2012ApJ...759...23S},
and then we have $f_{\rm IGM}\approx 0.82$. Indeed, the slowly evolving $f_{\rm IGM}$ at low redshifts is also a big
challenge for the cosmological applications of FRBs.

Recently, ref. \cite{2019ApJ...876..146L} proposed that a nearly cosmology-free estimation for the fraction of baryons in the IGM
can be achieved by using a putative sample of FRBs with the measurements of both DM and luminosity distance $d_{\rm L}$.
In principle, within the framework of the standard $\Lambda$CDM model, $f_{\rm IGM}$ can be directly estimated by using
the usual DM--$z$ way. However, the $\Lambda$CDM model currently faces the so-called Hubble constant tension problem,
which is the $4\sigma$ discrepancy between the expansion rate directly determined from local distance measurements \citep{2019ApJ...876...85R}
and the one obtained in the context of the $\Lambda$CDM model from high-redshift cosmic microwave background radiation data \citep{2018arXiv180706209P}.
Moreover, the $\Lambda$CDM model presents some puzzles for theorists, such as the nature of dark matter and dark energy
as well as fine-tuning and coincidence problems \citep{1989RvMP...61....1W,2000astro.ph..5265W}. In order to alleviate
these problems, numerous alternative models have been proposed. This scenario makes clear that cosmology-independent
measurements of cosmological quantities are fundamental importance for a proper evaluation of the possibilities.
In other words, any model-independent methods for estimating $f_{\rm IGM}$ are worthy of consideration.

In this paper, we propose a new geometric and cosmology-free method to constrain the evolution of $f_{\rm IGM}(z)$
by combining two independent measurements of the Hubble parameter $H(z)$ and the dispersion measure ${\rm DM}(z)$.
For $H(z)$ data, it can be obtained from differential ages of galaxies \cite{2002ApJ...573...37J,2005PhRvD..71l3001S,2010JCAP...02..008S}
and from radial baryon acoustic oscillation data \cite{2009MNRAS.399.1663G,2012MNRAS.425..405B,2013MNRAS.429.1514S}.
As for ${\rm DM}(z)$ data, we use the large FRB sample.
The rest of the paper is arranged as follows. In Section~\ref{sec:method}, we introduce the model-independent method
used to determine $f_{\rm IGM}(z)$. In Section~\ref{sec:simulation}, we test the validity and efficiency of our new method
using Monte Carlo simulations. Finally, a brief summary and discussion are drawn in Section~\ref{sec:summary}.

\section{Model-independent Method for Determining $f_{\rm IGM}(z)$}
\label{sec:method}

The observed DM of an FRB consists of several components:
\begin{equation}
{\rm DM_{obs}=DM_{MW}+DM_{IGM}}+\frac{\rm DM_{host}}{1+z}\;,
\label{eq:DM}
\end{equation}
where ${\rm DM_{MW}}$, ${\rm DM_{IGM}}$, and ${\rm DM_{host}}$ represent the DM contributions from the Milky Way, the IGM,
and the FRB host galaxy (including the host galaxy interstellar medium and the near-source plasma), respectively.
The cosmological redshift factor, $1+z$, converts the DM measured by the rest-frame observer to that of the Earth observer
\citep{2003ApJ...598L..79I,2014ApJ...783L..35D}. The average $\rm DM_{IGM}$ caused by the inhomogeneous IGM can be estimated as
%\begin{eqnarray}
%  {\rm DM}_{\rm IGM}(z)&=& \nonumber \frac{3cH_0\Omega_b }{8\pi G m_p}\times\\
%  & &\int_{0}^{z} \frac{(1+z') f_{\rm IGM}(z') \chi(z') dz'}{\sqrt{\Omega_{m} (1+z')^{3}+\left(1-\Omega_{m}\right)(1+z')^{3(1+w_0)}}}\;,
%\label{eq:IGM}
%\end{eqnarray}
\begin{equation}
  \langle{\rm DM}_{\rm IGM}\rangle(z)= \frac{3c\Omega_b H_0^{2}}{8\pi G m_p}
  \int_{0}^{z} \frac{(1+z') f_{\rm IGM}(z') \chi(z')}{H(z')} dz'\;,
\label{eq:IGM}
\end{equation}
where $\Omega_b$ is the present-day baryon density parameter of the universe, $f_{\rm IGM}(z)$ is the fraction of baryons in the IGM,
$H(z)=H_0[\Omega_{m} (1+z)^{3}+(1-\Omega_{m})(1+z)^{3(1+w_0)}]^{1/2}$ is the Hubble parameter at redshift $z$ in the $w$CDM model, and
\begin{equation}
\chi(z)=Y_{\rm H}\chi_{\rm e,H}(z)+\frac{1}{2}Y_{\rm He}\chi_{\rm e,He}(z)
\end{equation}
is the free electron number fraction per baryon, with $Y_{\rm H}=3/4$ and $Y_{\rm He}=1/4$ denoting the mass fractions of hydrogen and helium,
respectively, and $\chi_{\rm e,H}(z)$ and $\chi_{\rm e,He}(z)$ denoting the ionization fractions
of hydrogen and helium, respectively. As both hydrogen and helium are fully ionized at $z<3$
\citep{2006ARA&A..44..415F,2009ApJ...694..842M,2009RvMP...81.1405M,2011MNRAS.410.1096B},
it is reasonable to take $\chi_{\rm e,H}(z)=\chi_{\rm e,He}(z)=1$ for nearby FRBs. One then has $\chi(z)\simeq7/8$.
Different from previous studies \citep{2003ApJ...598L..79I,2004MNRAS.348..999I,2014ApJ...783L..35D},
we call Equation~(\ref{eq:IGM}) the ``average ${\rm DM_{IGM}}$'' since the IGM is considered to be inhomogeneous
and large fluctuation of individual line of sight is expected \citep{2014ApJ...780L..33M}.

In order to extract ${\rm DM_{IGM}}$ of an FRB, one needs to know ${\rm DM_{MW}}$ and ${\rm DM_{host}}$.
For a well-localized FRB, ${\rm DM_{MW}}$ can be well derived based on the Galactic electron density models
of ref. \cite{2002astro.ph..7156C} or ref. \cite{2017ApJ...835...29Y}. However, it is difficult to derive ${\rm DM_{host}}$
from an individual FRB, because it depends on the type of the host galaxy, the relative orientations of the FRB host
and source, and the near-source plasma \citep{2015RAA....15.1629X}. Ref. \cite{2018MNRAS.481.2320L} modeled
the FRB host DM contribution as a function of redshift by assuming that the rest-frame ${\rm DM_{host}}$
distribution accommodates the evolution of star formation history, i.e.,
${\rm DM_{host}}(z)={\rm DM_{host,0}}\sqrt{\frac{{\rm SFR}(z)}{{\rm SFR}(0)}}$, where ${\rm DM_{host,0}}$
is the present value of ${\rm DM_{host}}(z=0)$ and ${\rm SFR}(z)=\frac{0.0157+0.118z}{1+(z/3.23)^{4.66}}$
$\rm M_{\odot}$ $\rm yr^{-1}$ $\rm Mpc^{-3}$ is the adopted star formation rate \citep{2006ApJ...651..142H,2008MNRAS.388.1487L}.
In addition, while ${\rm DM_{IGM}}$ increases with redshift, ${\rm DM_{host}}$ becomes less significant
at high redshifts due to the $(1+z)$ factor. Therefore, we can roughly subtract ${\rm DM_{host}}$ from ${\rm DM_{obs}}$
and leave its uncertainty $(\sigma_{\rm host}=\sigma_{\rm host,0}\sqrt{\frac{{\rm SFR}(z)}{{\rm SFR}(0)}})$
into the total uncertainty $\sigma_{\rm tot}$, which is the uncertainty of ${\rm DM_{IGM}}$ extracted from ${\rm DM_{obs}}$.
It has
\begin{equation}
\sigma_{\rm tot} = \left[\sigma_{\rm obs}^{2}+\sigma_{\rm MW}^{2}+\sigma_{\rm IGM}^{2}
+\left(\frac{\sigma_{\rm host}}{1+z}\right)^{2} \right]^{1/2}\;.
\label{eq:sigmaIGM}
\end{equation}
Based on current observations compiled in the FRB catalog (see ref. \cite{2016PASA...33...45P} and references therein),
we adopt an average value $\sigma_{\rm obs}=1.5$ pc $\rm cm^{-3}$ as the uncertainty of $\rm DM_{obs}$.
For high Galactic latitude ($|b|>10^{\circ}$) sources, the average uncertainty of $\rm DM_{MW}$ is about
$10$ pc $\rm cm^{-3}$ \citep{2005AJ....129.1993M}. As pointed out in ref. \cite{2014ApJ...780L..33M},
for individual FRBs, the detected ${\rm DM_{IGM}}(z)$ may deviate significantly from $\langle{\rm DM}_{\rm IGM}\rangle(z)$ given in Equation~(\ref{eq:IGM}).
The sightline-to-sightline scatter in ${\rm DM_{IGM}}(z)$ is related to the profile models characterizing the inhomogeneity
of the baryon matter in the IGM (see Figure~1 of ref. \cite{2014ApJ...780L..33M}). Here we associate the standard deviation $\sigma_{\rm IGM}(z)$ derived from the simulations
of ref. \cite{2014ApJ...780L..33M} to ${\rm DM_{IGM}}(z)$. Due to the effects of inhomogeneities, the ratio $\sigma_{\rm IGM}/\langle{\rm DM}_{\rm IGM}\rangle$
is very large. Fortunately, if there are enough FRBs from different sightlines but in a narrow redshift bin, their weighted average $\overline{\rm DM}_{\rm IGM}(z)$
would be a reasonable approximation of $\langle{\rm DM}_{\rm IGM}\rangle(z)$ \citep{2014PhRvD..89j7303Z}. We thus use the data-binning procedure to estimate
the derivative ${\rm DM'_{IGM}}$ (more on this below).
Following ref. \cite{2019ApJ...876..146L}, we take $\sigma_{\rm host,0}=30$
pc $\rm cm^{-3}$ as the uncertainty of ${\rm DM_{host,0}}$.

Differentiating Equation~(\ref{eq:IGM}), we can obtain an expression for the fraction of baryon mass in the IGM:
\begin{equation}
f_{\rm IGM}(z) = \frac{H(z){\rm DM}'_{\rm IGM}(z)}{A(1+z)}\;,
\label{eq:fIGM}
\end{equation}
where $A=\frac{21c\Omega_b H_0^{2}}{64\pi G m_p}$ and ${\rm DM}'_{\rm IGM}(z)=d{\rm DM_{IGM}}(z)/dz$
denotes the first derivative with respective to redshift $z$. As suggested in Equation~(\ref{eq:fIGM}) that
the baryon fraction in the IGM, $f_{\rm IGM}(z)$, can be cosmological-model-independently determined
by combining the Hubble expansion rate $H(z)$ and DM measurements. Remarkably, this estimation achieves
the evolution of $f_{\rm IGM}(z)$ from these measurements at any single redshift.

\section{Simulations and Results}
\label{sec:simulation}
In this section, we perform Monte Carlo simulations to test the efficiency of our new model-independent method.
Here we adopt the fiducial flat $\Lambda$CDM model with the cosmological parameters derived from the latest \emph{Planck} data
($H_{0}=67.36$ km $\rm s^{-1}$ $\rm Mpc^{-1}$, $\Omega_{m}=0.315$, and $\Omega_b=0.0493$) \cite{2018arXiv180706209P}.
In order to determine the baryon fraction in the IGM, $f_{\rm IGM}$, at different redshifts, one needs to combine two independent measurements
of the Hubble parameter $H(z)$ and the dispersion measure ${\rm DM_{IGM}}(z)$ (or the derivative of ${\rm DM_{IGM}}(z)$).

\begin{figure}
\vskip-0.3in
\centerline{\includegraphics[angle=0,width=1.0\hsize]{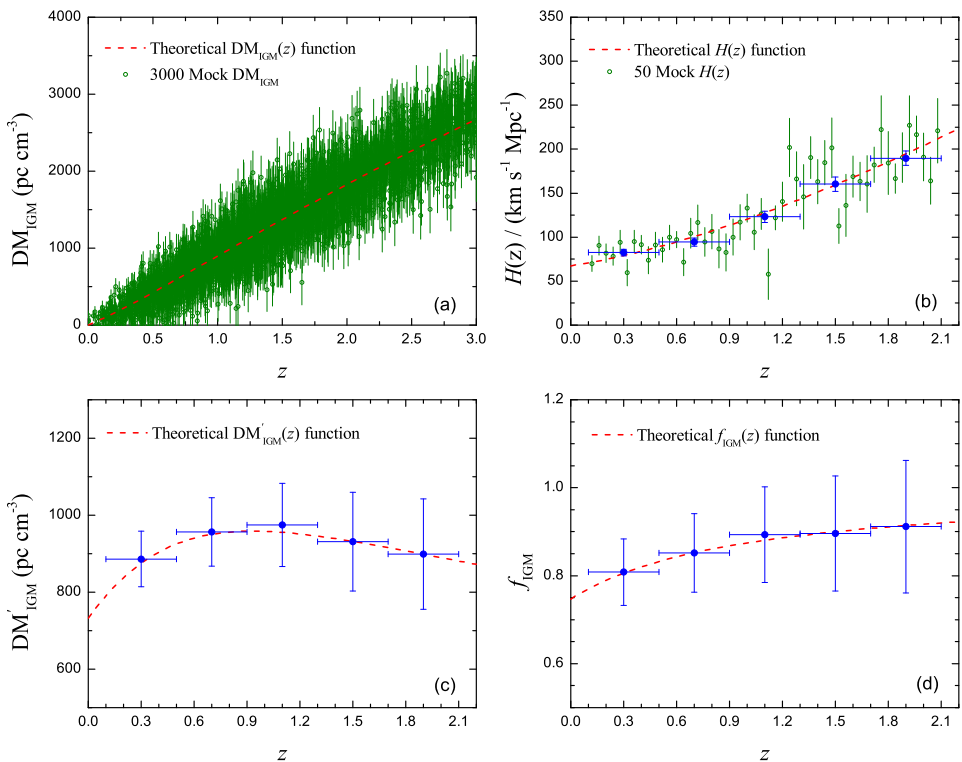}}
\vskip-0.4in
\caption{Example of the simulations for the case of 3000 simulated FRBs and 50 simulated $H(z)$ data.
Panel (a) shows 3000 mock ${\rm DM_{IGM}}$ data (green circles) and the theoretical ${\rm DM_{IGM}}(z)$ function (dashed line).
Panel (b) shows 50 mock $H(z)$ data (green circles), five binned mock $H(z)$ data (blue dots), and the theoretical $H(z)$ function (dashed line).
Panels (c) and (d) show the final constraints on ${\rm DM}'_{\rm IGM}$ and $f_{\rm IGM}$ at different redshifts (blue dots)
derived from these mock data with the binning method, respectively. The theoretical ${\rm DM}'_{\rm IGM}(z)$ and $f_{\rm IGM}(z)$
functions (dashed lines) are also displayed for comparison. Horizontal error bars in Panels (b), (c), and (d) correspond to
the ranges of each redshift bin.}
\label{f1}
\end{figure}

Firstly, we simulate ${\rm DM_{IGM}}(z)$ data of future FRBs. In previous studies, $f_{\rm IGM}$ is often assumed to be a constant.
But in reality, as massive halos become less abundant in the early universe, $f_{\rm IGM}$ should grow with redshift
\citep{2014ApJ...780L..33M}. In our simulations, we parameterize $f_{\rm IGM}$ as a mildly increasing function of redshift,
$f_{\rm IGM}(z)=f_{\rm IGM,0}+\alpha z/(1+z)$, as ref. \cite{2019ApJ...876..146L} did in their treatment.
The estimated values of $f_{\rm IGM}$ are 0.82 and 0.9 at $z\leq0.4$ and $z\geq1.5$, respectively \citep{2009RvMP...81.1405M,2012ApJ...759...23S}.
Substituting $f_{\rm IGM}(z=0.4)=0.82$ and $f_{\rm IGM}(z=1.5)=0.9$ into the parameterized $f_{\rm IGM}(z)$ function,
one then has $f_{\rm IGM}(z)=0.747+0.255 z/(1+z)$. Thus, we take $f_{\rm IGM,0}=0.747$ and $\alpha=0.255$ as the fiducial
values for DM simulations. The redshift distribution of FRBs is assumed as $P(z)\propto z e^{-z}$ in the redshift range
$0<z<3$, which is a phenomenological model for the redshift distribution of gamma-ray bursts (GRBs) \citep{2011ApJ...738...19S,2014PhRvD..89j7303Z}.
With the mock $z$, we infer the fiducial value of $\rm DM_{IGM}^{fid}$ from Equation~(\ref{eq:IGM}).
We then add the deviation $\sigma_{\rm tot}$ in Equation~(\ref{eq:sigmaIGM}) to the fiducial value of $\rm DM_{IGM}^{fid}$.
That is, the simulated DM, $\rm DM_{IGM}^{sim}$, is sampled from the normal distribution
${\rm DM_{IGM}^{sim}}=\mathcal{N}({\rm DM_{IGM}^{fid}},\;\sigma_{\rm tot})$.
Thanks to the high event rate ($\sim10^{4}$ $\rm sky^{-1}$ $\rm day^{-1}$) \cite{2013Sci...341...53T,2016MNRAS.460L..30C},
FRBs are expected to be detected in the tens of thousands by the upcoming Canadian Hydrogen Intensity Mapping Experiment (CHIME)
\cite{2018ApJ...863...48C} and Hydrogen Intensity and Real-time Analysis experiment (HIRAX) \cite{2016SPIE.9906E..5XN} radio arrays.
We first run a simulation with 3000 mock FRBs probably detected with CHIME or HIRAX in a few years.
Panel (a) of Figure~\ref{f1} shows an example of 3000 simulated ${\rm DM_{IGM}}(z)$ data.

By the time we have a few thousands of FRBs, there might be more $H(z)$ measurements from different observables,
such as the Baryon Oscillation Spectroscopic Survey\footnote{http://www.sdss3.org/surveys/boss.php}, the proposed
Sandage-Loeb observational plan \citep{2007PhRvD..75f2001C,2010PhRvD..82l3513A,2010PhLB..691...11Z},
and the Atacama Cosmology Telescope\footnote{http://www.physics.princeton.edu/act/index.html.}. Especially,
the Atacama Cosmology Telescope might be able to identify $\sim2000$ passively evolving galaxies up to $z\approx1.5$
via the Sunyaev-Zel'dovich effect, and their spectra could be analyzed to yield age determinations that would obtain
$\sim1000$ $H(z)$ measurements \citep{2005PhRvD..71l3001S}. But it is difficult to figure out how many
$H(z)$ measurements will actually be yielded by the time that 3000 FRBs are detected. Since the number of currently
observational $H(z)$ data is around 40 (see \cite{2017ApJ...838..160W} and references therein), we conservatively
assume that there will be 50 $H(z)$ data points when we have 3000 FRBs, the redshifts of which are chosen equally
within the range $0.1\leq z\leq2.1$.
The uncertainties of current $H(z)$ measurements are
in the region confined by two straight lines $\sigma_{+}=16.87z+10.48$ and $\sigma_{-}=4.41z+7.25$ from above
to below, respectively \citep{2011ApJ...730...74M}. If we believe that future observations of $H(z)$ would also
have uncertainties at this level, we can draw their uncertainties $\sigma_{H}(z)$ from the Gaussian distribution
$\mathcal{N}(\sigma_{0}(z),\;\epsilon(z))$, where $\sigma_{0}(z)=(\sigma_{+}+\sigma_{-})/2$ is the mean uncertainty
and $\epsilon(z)=(\sigma_{+}-\sigma_{-})/4$ is chosen so that the probability of $\sigma_{H}(z)$ falling within
the confined area is 95.4\% \citep{2011ApJ...730...74M}. With the fiducial value of $H^{\rm fid}(z)$ inferred from
the flat $\Lambda$CDM model, we sample the simulated Hubble parameter, $H^{\rm sim}(z)$, according to the Gaussian
distribution $H^{\rm sim}(z)=\mathcal{N}(H^{\rm fid}(z),\;\sigma_{H}(z))$. An example of 50 simulated $H(z)$ data
(green circles) is presented in Figure~\ref{f1}(b).

\begin{table}
%\small
\footnotesize
\centering \caption{Model-independent Estimates of the Fraction of Baryon Mass in the IGM from FRB and $H(z)$ data}
\begin{tabular}{lcccccc}
\hline
\hline
  &   &   &   & \multicolumn{1}{c}{3,000 FRBs + 50 $H(z)$} & \multicolumn{1}{c}{6,000 FRBs + 100 $H(z)$}  & \multicolumn{1}{c}{12,000 FRBs + 200 $H(z)$}  \\
\cline{5-7}
Bins  &  $z_{\rm min}$  &  $z_{\rm max}$   &  $\bar{z}$  &  $f_{\rm IGM}(z)$  &  $f_{\rm IGM}(z)$  &  $f_{\rm IGM}(z)$ \\
\hline
1  &  0.1  &  0.5  &  0.3   &   $0.808\pm0.076$   &   $0.808\pm0.054$   &   $0.807\pm0.038$   \\
2  &  0.5  &  0.9  &  0.7   &   $0.852\pm0.089$   &   $0.852\pm0.064$   &   $0.851\pm0.045$  \\
3  &  0.9  &  1.3  &  1.1   &   $0.893\pm0.109$   &   $0.892\pm0.077$   &   $0.891\pm0.054$   \\
4  &  1.3  &  1.7  &  1.5   &   $0.896\pm0.131$   &   $0.896\pm0.091$   &   $0.896\pm0.064$   \\
5  &  1.7  &  2.1  &  1.9   &   $0.911\pm0.151$   &   $0.911\pm0.107$   &   $0.911\pm0.076$   \\
\hline
\multicolumn{4}{c}{$\langle \sigma_{f_{\rm IGM}}/f_{\rm IGM} \rangle$} & 12.6\%  & 8.9\%  &  6.3\% \\
\hline
\end{tabular}
\label{table1}
\medskip \\
$^{\rm Note.}$$z_{\rm min}$ and $z_{\rm max}$  correspond to the left and right edges of each redshift bin, respectively.
$\bar{z}$ represents the mean redshift of the bin.
\end{table}

The remaining issue is to reasonably estimate the derivative of ${\rm DM_{IGM}}(z)$ with respect to $z$, ${\rm DM}'_{\rm IGM}(z)$,
from the mock FRB data. In order to maximize the use of all available mock data, we take the following data-binning procedure
to estimate ${\rm DM}'_{\rm IGM}(z)$ and then to determine $f_{\rm IGM}(z)$ \citep{2014PhRvD..90b3012S}. We split the redshift interval into five bins
delimited by the following redshifts: $z=\{0.1,0.5,0.9,1.3,1.7,2.1\}$. This choice was made for two main reasons.
First, such a large uncertainty of ${\rm DM_{IGM}}(z)$, $\sigma_{\rm tot}$ (see Figure~\ref{f1}(a)), is a serious challenge for
using FRBs as a cosmic probe, but it is feasible by using the weighted average ${\rm \overline{DM}_{IGM}}(z)$ when there are sufficient FRBs
in a narrow redshift bin. Second, we do not want to have redshift bins that are too large, in order for the assumption that
${\rm DM_{IGM}}(z)$ and $H(z)$ be constant inside each bin to be still reasonable. The resulting bins are listed in Table~\ref{table1}.
In each bin, we then calculate the weighted average of all available $H(z)$ through
\begin{equation}
%{\rm \overline{DM}_{IGM}}=\frac{\sum_{i}{\rm DM}_{{\rm IGM},i}/\sigma^{2}_{{\rm tot},i}}{\sum_{i}1/\sigma^{2}_{{\rm tot},i}}\;{\rm and}\;
\overline{H}(z)=\frac{\sum_{i}H(z_{i})/\sigma^{2}_{H,i}}{\sum_{i}1/\sigma^{2}_{H,i}}\;.
\end{equation}
The corresponding uncertainty on $\overline{H}(z)$ can be obtained from
\begin{equation}
%\sigma^{2}_{\rm \overline{DM}_{IGM}}=\frac{1}{\sum_{i}1/\sigma^{2}_{{\rm tot},i}}\;{\rm and}\;
\sigma^{2}_{\overline{H}(z)}=\frac{1}{\sum_{i}1/\sigma^{2}_{H,i}}\;.
\end{equation}
We assign these values to the mean redshifts of different $H(z)$ that contained in each bin, i.e., $\overline{z}_H=\{0.3,0.7,1.1,1.5,1.9\}$.
In Figure~\ref{f1}(b) we show $\overline{H}(\overline{z}_H)$ (blue dots) obtained with this binning.
To compute the derivative ${\rm DM}'_{\rm IGM}(z)$, we adopt the following formula as its discrete approximation:
\begin{equation}
{\rm DM}'_{\rm IGM}(z)\simeq \frac{{\rm DM_{IGM}}(z+\Delta z_{1})-{\rm DM_{IGM}}(z+\Delta z_{2})}{\Delta z_{1}+\Delta z_{2}}\;.
\end{equation}
In order to apply this formula to our FRB data, we divide each bin into two equal sub-bins, taking
$\overline{z}_H$ as the splitting point. We then compute the weighted average ${\rm \overline{DM}_{IGM}}$ in each sub-bin,
obtaining ${\rm \overline{DM}_{left}}$ and ${\rm \overline{DM}_{right}}$. We assign ${\rm \overline{DM}_{left}}$ and ${\rm \overline{DM}_{right}}$
to the average redshifts of the corresponding sub-bins, $\overline{z}_{\rm left}$ and $\overline{z}_{\rm right}$,
and finally compute the approximated derivative as
\begin{equation}
{\rm DM}'_{\rm IGM}(\overline{z}_H)=\frac{{\rm \overline{DM}_{right}}-{\rm \overline{DM}_{left}}}{\overline{z}_{\rm right}-\overline{z}_{\rm left}}\;.
\end{equation}

\begin{figure}
\vskip-0.3in
\centerline{\includegraphics[keepaspectratio,clip,width=1.0\textwidth]{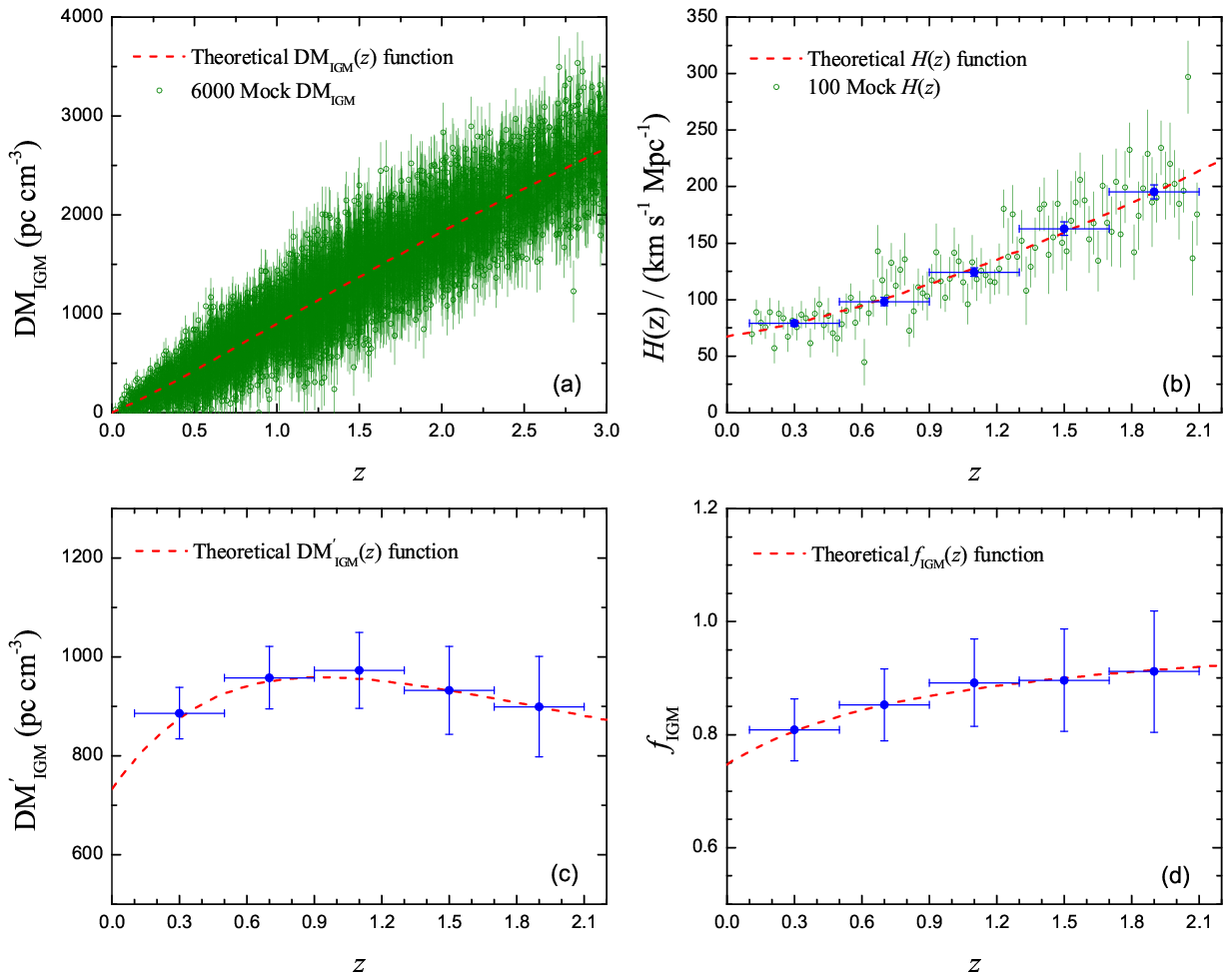}}
\vskip-0.4in
\caption{Same as Figure~\ref{f1}, but for the case of 6000 simulated FRBs and 100 simulated $H(z)$ data.}
\label{f2}
\end{figure}

\begin{figure}
\vskip-0.3in
\centerline{\includegraphics[keepaspectratio,clip,width=1.0\textwidth]{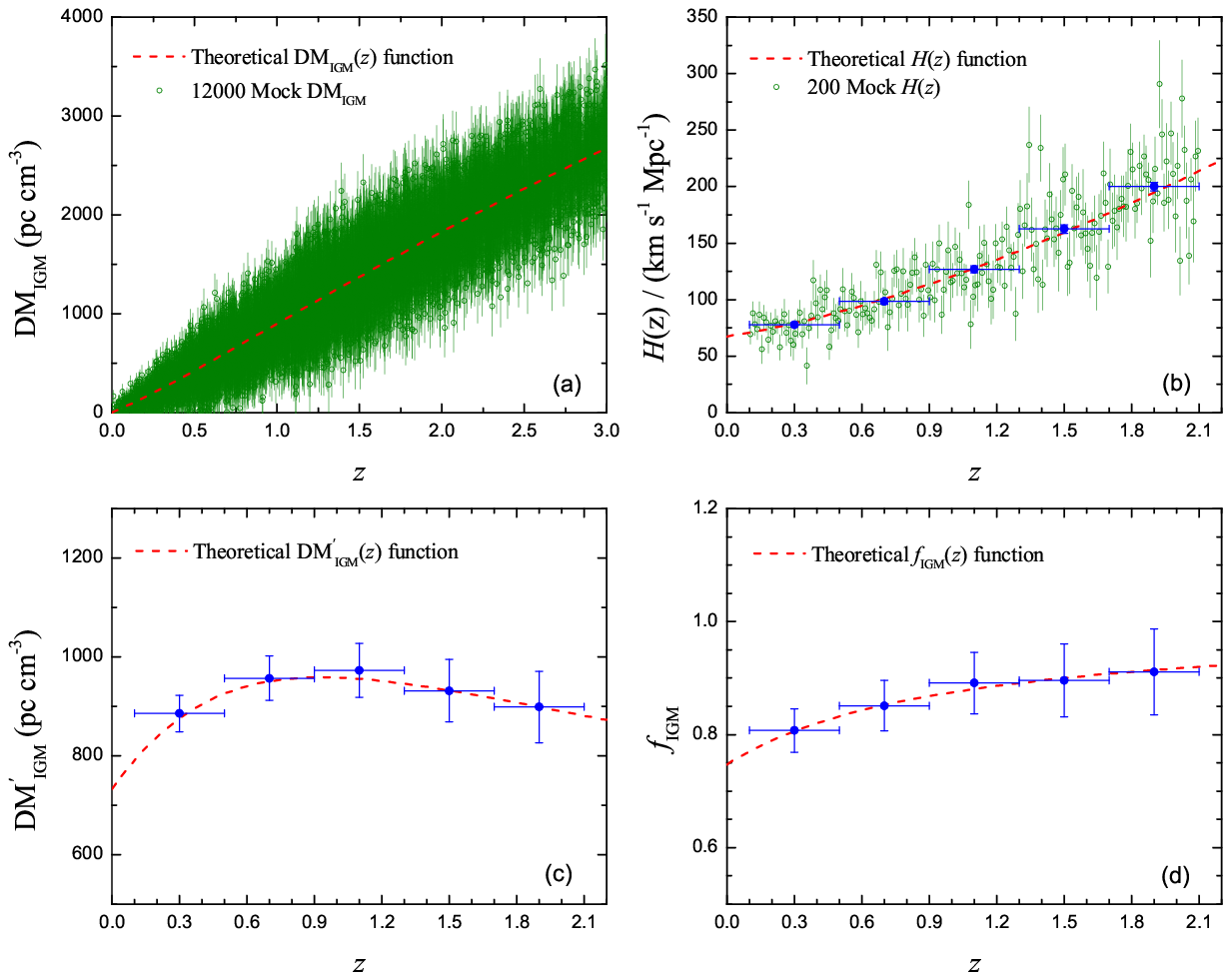}}
\vskip-0.4in
\caption{Same as Figure~\ref{f1}, but for the case of 12000 simulated FRBs and 200 simulated $H(z)$ data.}
\label{f3}
\end{figure}

Having obtained $\overline{H}(\overline{z}_H)$ and ${\rm DM}'_{\rm IGM}(\overline{z}_H)$, we can finally compute $f_{\rm IGM}(\overline{z}_H)$
at each redshift $\overline{z}_H$ through Equation~(\ref{eq:fIGM}), where, as already said, ${\rm DM}'_{\rm IGM}(\overline{z}_H)$
are computed from FRB data as explained above and $\overline{H}(\overline{z}_H)$ come from Hubble parameter measurements.
To avoid the randomness of simulation, we repeat the simulation process 1000 times for each FRB $+$ $H(z)$ data sets and
draw the means of ${\rm DM}'_{\rm IGM}$ and $f_{\rm IGM}$ as well as their corresponding root mean square of uncertainties.
The final constraints on ${\rm DM}'_{\rm IGM}$ and $f_{\rm IGM}$ (blue dots) together with their $1\sigma$ error bars
are displayed in panels (c) and (d) of Figure~\ref{f1}, respectively.
Because of the poor quality of mock data at higher redshifts, the errors become larger. For a comparison, we also give the
theoretical ${\rm DM}'_{\rm IGM}(z)$ and $f_{\rm IGM}(z)$ functions (dashed lines) in panels (c) and (d) of Figure~\ref{f1}, respectively.
It is found that the final derived ${\rm DM}'_{\rm IGM}$ and $f_{\rm IGM}$ data (blue dots) are well consistent with
theoretical functions (dashed lines). This suggests that the discrete ${\rm DM}'_{\rm IGM}$ and $f_{\rm IGM}$ data derived
from mock FRB $+$ $H(z)$ data with the binning method are reliable. Additionally, we find that, from 3000 simulated FRBs and 50
simulated $H(z)$ data points, the unbiased constraints on the fraction of baryon mass in IGM at different redshifts are
$f_{\rm IGM}(z=0.3)=0.808\pm0.076$, $f_{\rm IGM}(z=0.7)=0.852\pm0.089$, $f_{\rm IGM}(z=1.1)=0.893\pm0.109$, $f_{\rm IGM}(z=1.5)=0.896\pm0.131$,
and $f_{\rm IGM}(z=1.9)=0.911\pm0.151$, respectively. The mean relative error of these five $f_{\rm IGM}$ data is $\langle \sigma_{f_{\rm IGM}}/f_{\rm IGM} \rangle \simeq12.6\%$.
Note that as the relative errors of ${\rm DM}'_{\rm IGM}(\overline{z}_H)$ are much larger than those of $H(\overline{z}_H)$ at the same redshifts
(see Figures~\ref{f1}(b) and (c)), the precision of the derived $f_{\rm IGM}(\overline{z}_H)$ is strongly dominated by the uncertainty of
${\rm DM}'_{\rm IGM}(\overline{z}_H)$. That is, the chosen number of $H(z)$ measurements has little impact on our results.

At this point, it is interesting to compare our forecast result with the other model-independent method.
Ref. \cite{2019ApJ...876..146L} showed that a cosmology-independent estimate of the value of $f_{\rm IGM}$ at $z=1.5$
with a $\sim14.0\%$ uncertainty can be obtained by using 500 FRBs with the measurements of both DM and $d_{\rm L}$
(i.e., the $d_{\rm L}/$DM method). The relative uncertainty on the determined $f_{\rm IGM}(z=1.5)$ is at the level
of $\sigma_{f_{\rm IGM}}/f_{\rm IGM}\simeq14.6\%$ using 3000 FRBs with our new method, which is almost as well as
that of using 500 FRBs with the $d_{\rm L}/$DM method. Though our method needs six times the number of FRBs
to achieve the same precision, the joint measurements of DM and $d_{\rm L}$ suggested by the $d_{\rm L}/$DM method
may not be easy in practice. It is not clear what fraction of FRBs will actually satisfy the joint measurements.
Most importantly, our method provides a new cosmology-independent way to constrain $f_{\rm IGM}(z)$.

To better represent how effective our method might be with more FRB and $H(z)$ measurements, we also carry out a similar
analysis by considering the case of 6000 simulated FRBs and 100 simulated $H(z)$ data. Simulations for ${\rm DM_{IGM}}$
and $H(z)$ are shown in panels (a) and (b) of Figure~\ref{f2}, respectively. The final constraints on ${\rm DM}'_{\rm IGM}$ and
$f_{\rm IGM}$ at different redshifts derived from these mock data with the binning method are presented in
panels (c) and (d) of Figure~\ref{f2}, respectively. The mean relative error of $f_{\rm IGM}$ in five bins is $\sim8.9\%$.
Finally, we investigate the case of 12000 simulated FRBs and 200 simulated $H(z)$ data. Simulations for ${\rm DM_{IGM}}$
and $H(z)$ are shown in panels (a) and (b) of Figure~\ref{f3}, respectively. Corresponding constraints on ${\rm DM}'_{\rm IGM}$ and
$f_{\rm IGM}$ are presented in panels (c) and (d) of Figure~\ref{f3}, respectively. The mean relative error of $f_{\rm IGM}$ is $\sim6.3\%$.
The resulting constraints on the evolution of $f_{\rm IGM}$ for different FRB + $H(z)$ cases are summarized in Table~\ref{table1}.
As expected, the uncertainty of $f_{\rm IGM}$ is gradually reduced with the increasing number of FRBs and $H(z)$. These
results imply that the evolution of $f_{\rm IGM}$ can be unbiasedly inferred in a model-independent manner when
$\sim\mathcal{O}(10^{3})$ FRBs are detected.

\begin{figure}
\vskip-0.2in
\centerline{\includegraphics[keepaspectratio,clip,width=1.0\textwidth]{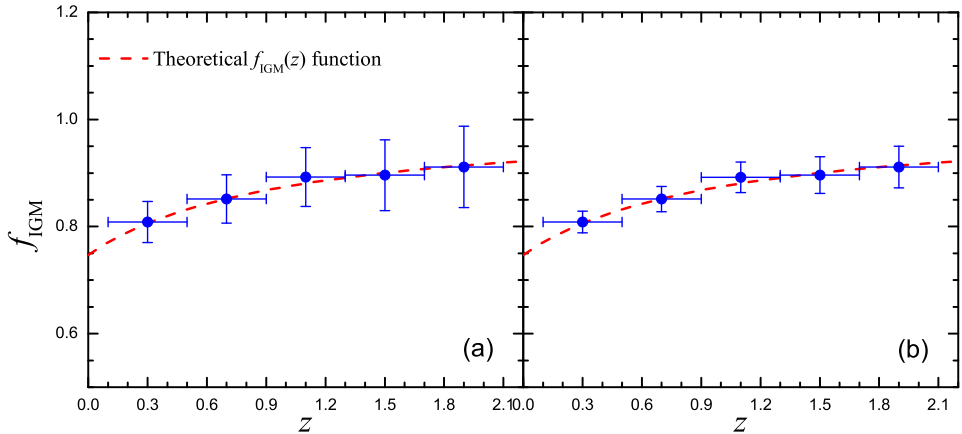}}
\vskip-0.2in
\caption{$f_{\rm IGM}(z)$ estimations obtained from the mock data sets of 3000 FRBs and 50 $H(z)$,
but with a half of (left panel) and a quarter of (right panel) the original errors, respectively.}
\label{f4}
\end{figure}

To understand how the quality of simulated data affects the constraints, we perform parallel comparative analyses of
the mock data sets by reducing the original errors by a factor of 0.5 and 0.25, respectively. We also assume that
$N_{\rm FRB}=3000$ and $N_{H(z)}=50$. For the analysis with a half of the errors, as shown in Table~\ref{table2}
and the left panel of Figure~\ref{f4}, we obtain $\langle \sigma_{f_{\rm IGM}}/f_{\rm IGM} \rangle \simeq6.4\%$.
For the case with a quarter of the errors, as shown in Table~\ref{table2} and the right panel of Figure~\ref{f4}, we obtain
$\langle \sigma_{f_{\rm IGM}}/f_{\rm IGM} \rangle \simeq3.3\%$. One can see that the precision of $f_{\rm IGM}$
can be significantly improved with the increasing quality of FRBs and $H(z)$. Additionally, increasing the quality
of future FRB and $H(z)$ data is more important than increasing their quantity.

\begin{table}
%\small
\footnotesize
\centering \caption{$f_{\rm IGM}(z)$ estimations obtained from the mock data sets of 3000 FRBs and 50 $H(z)$ but with an increase in their quality}
\begin{tabular}{lccccc}
\hline
\hline
  &   &   &   & \multicolumn{1}{c}{with Errors Reduced by 50\%} & \multicolumn{1}{c}{with Errors Reduced by 75\%}  \\
\cline{5-6}
Bins  &  $z_{\rm min}$  &  $z_{\rm max}$   &  $\bar{z}$  &  $f_{\rm IGM}(z)$  &  $f_{\rm IGM}(z)$  \\
\hline
1  &  0.1  &  0.5  &  0.3   &   $0.808\pm0.038$   &   $0.808\pm0.020$      \\
2  &  0.5  &  0.9  &  0.7   &   $0.851\pm0.045$   &   $0.851\pm0.024$     \\
3  &  0.9  &  1.3  &  1.1   &   $0.892\pm0.055$   &   $0.892\pm0.029$     \\
4  &  1.3  &  1.7  &  1.5   &   $0.896\pm0.066$   &   $0.896\pm0.034$      \\
5  &  1.7  &  2.1  &  1.9   &   $0.911\pm0.076$   &   $0.911\pm0.039$      \\
\hline
\multicolumn{4}{c}{$\langle \sigma_{f_{\rm IGM}}/f_{\rm IGM} \rangle$} & 6.4\%  & 3.3\%   \\
\hline
\end{tabular}
\label{table2}
\end{table}

\begin{figure}
\vskip-0.3in
\centerline{\includegraphics[keepaspectratio,clip,width=1.0\textwidth]{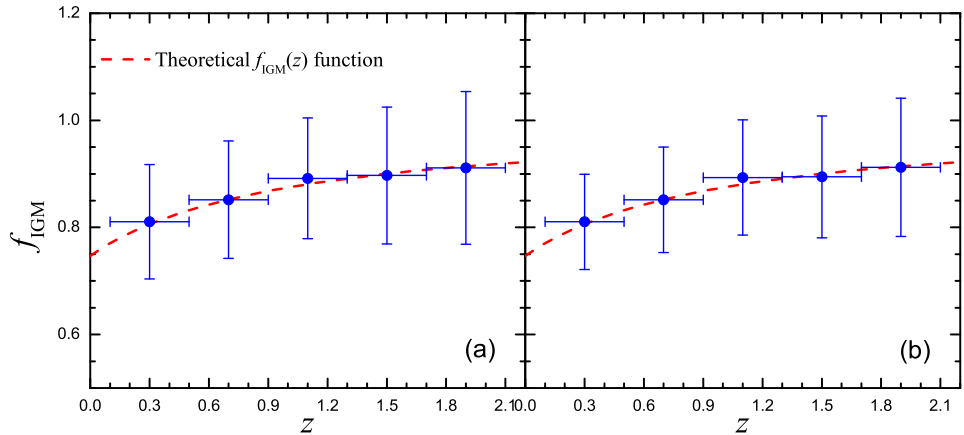}}
\vskip-0.2in
\caption{$f_{\rm IGM}(z)$ estimations obtained from the mock data sets of 3000 FRBs and 50 $H(z)$,
but with the redshift distributions of FRBs tracing star formation (left panel) and compact star mergers (right panel), respectively.}
\label{f5}
\end{figure}

In all the above simulations, we assume that the redshift distribution of FRBs satisfies $P(z)\propto z e^{-z}$
\citep{2011ApJ...738...19S,2014PhRvD..89j7303Z}, a phenomenological model for GRB redshift distribution.
While the redshift distribution of $H(z)$ measurements is assumed to be uniform. Since the precision of the
derived $f_{\rm IGM}(z)$ is mainly dominated by the uncertainty of ${\rm DM}'_{\rm IGM}(z)$, different $z$
distributions of $H(z)$ measurements would not affect the scatter of $f_{\rm IGM}(z)$, but FRB redshift
distribution might be. The true $z$ distribution of FRBs depends on the corresponding progenitor system(s) of FRBs,
which can take different empirical formulae (e.g., for the $z$ distributions tracing the history of star formation
or mergers of double compact stars, see approximate analytical forms in ref. \cite{2015ApJ...812...33S}).
In order to test how the $z$ distribution of FRBs affects our results, we perform simulations with two
different progenitor models. We also assume that $N_{\rm FRB}=3000$ and $N_{H(z)}=50$ with the original errors.
For the $z$ distribution tracing star formation, we adopt the approximate analytical model
derived by ref. \cite{2008ApJ...683L...5Y}: $P(z)=[(1+z)^{3.4\eta}+(\frac{1+z}{5000})^{-0.3\eta}+(\frac{1+z}{9})^{-3.5\eta}]^{1/\eta}$,
where $\eta=-10$. As shown in Table~\ref{table3} and the left panel of Figure~\ref{f5}, now we obtain
$\langle \sigma_{f_{\rm IGM}}/f_{\rm IGM} \rangle \simeq13.7\%$. For the $z$ distribution tracing compact star mergers,
we adopt the analytical model $P(z)=[(1+z)^{5.0\eta}+(\frac{1+z}{0.17})^{0.87\eta}+(\frac{1+z}{4.12})^{-8.0\eta}+(\frac{1+z}{4.05})^{-20.5\eta}]^{1/\eta}$,
where $\eta=-2$ \cite{2011ApJ...727..109V}. As shown in Table~\ref{table3} and the right panel of Figure~\ref{f5}, now we obtain
$\langle \sigma_{f_{\rm IGM}}/f_{\rm IGM} \rangle \simeq12.3\%$.
By comparing these constraints with those obtained from the phenomenological GRB model (see Figure~\ref{f1}(d)),
we can conclude that the explicit form of $z$ distribution does not affect the global shape of the $f_{\rm IGM}$--$z$ plot
we are modeling, but affects the scatter of $f_{\rm IGM}(z)$ to some extent.

\begin{table}
%\small
\footnotesize
\centering \caption{$f_{\rm IGM}(z)$ estimations obtained from the mock data sets of 3000 FRBs and 50 $H(z)$ but with different redshift distributions of FRBs}
\begin{tabular}{lccccc}
\hline
\hline
  &   &   &   & \multicolumn{1}{c}{star formation} & \multicolumn{1}{c}{compact star mergers}  \\
\cline{5-6}
Bins  &  $z_{\rm min}$  &  $z_{\rm max}$   &  $\bar{z}$  &  $f_{\rm IGM}(z)$  &  $f_{\rm IGM}(z)$  \\
\hline
1  &  0.1  &  0.5  &  0.3   &   $0.811\pm0.107$   &   $0.810\pm0.089$      \\
2  &  0.5  &  0.9  &  0.7   &   $0.852\pm0.109$   &   $0.852\pm0.098$     \\
3  &  0.9  &  1.3  &  1.1   &   $0.892\pm0.113$   &   $0.893\pm0.107$     \\
4  &  1.3  &  1.7  &  1.5   &   $0.897\pm0.128$   &   $0.895\pm0.114$      \\
5  &  1.7  &  2.1  &  1.9   &   $0.911\pm0.143$   &   $0.912\pm0.129$      \\
\hline
\multicolumn{4}{c}{$\langle \sigma_{f_{\rm IGM}}/f_{\rm IGM} \rangle$} & 13.7\%  & 12.3\%   \\
\hline
\end{tabular}
\label{table3}
\end{table}

\begin{figure}
%\vskip-0.1in
\includegraphics[keepaspectratio,clip,width=0.55\textwidth]{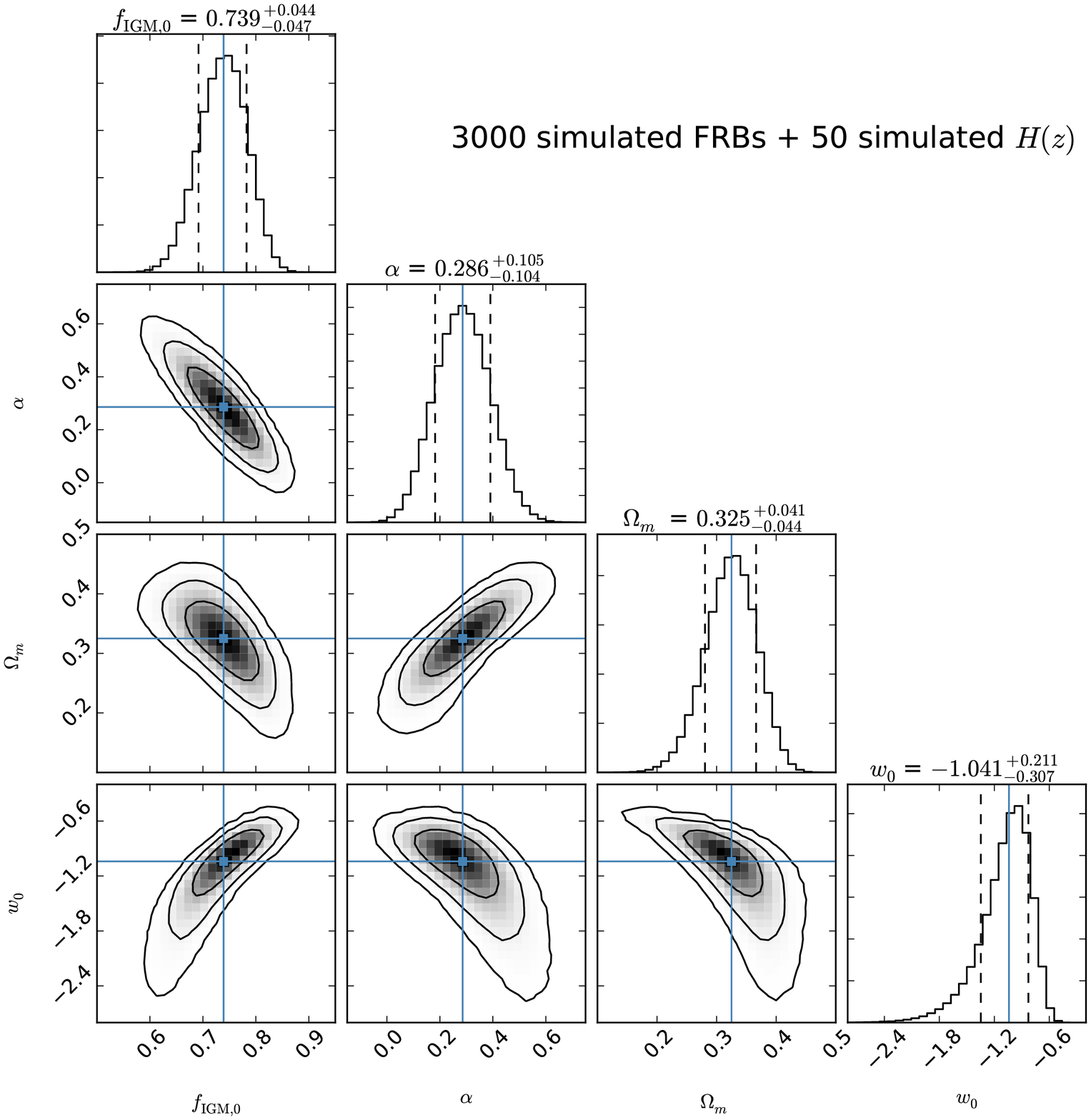}
\includegraphics[keepaspectratio,clip,width=0.5\textwidth]{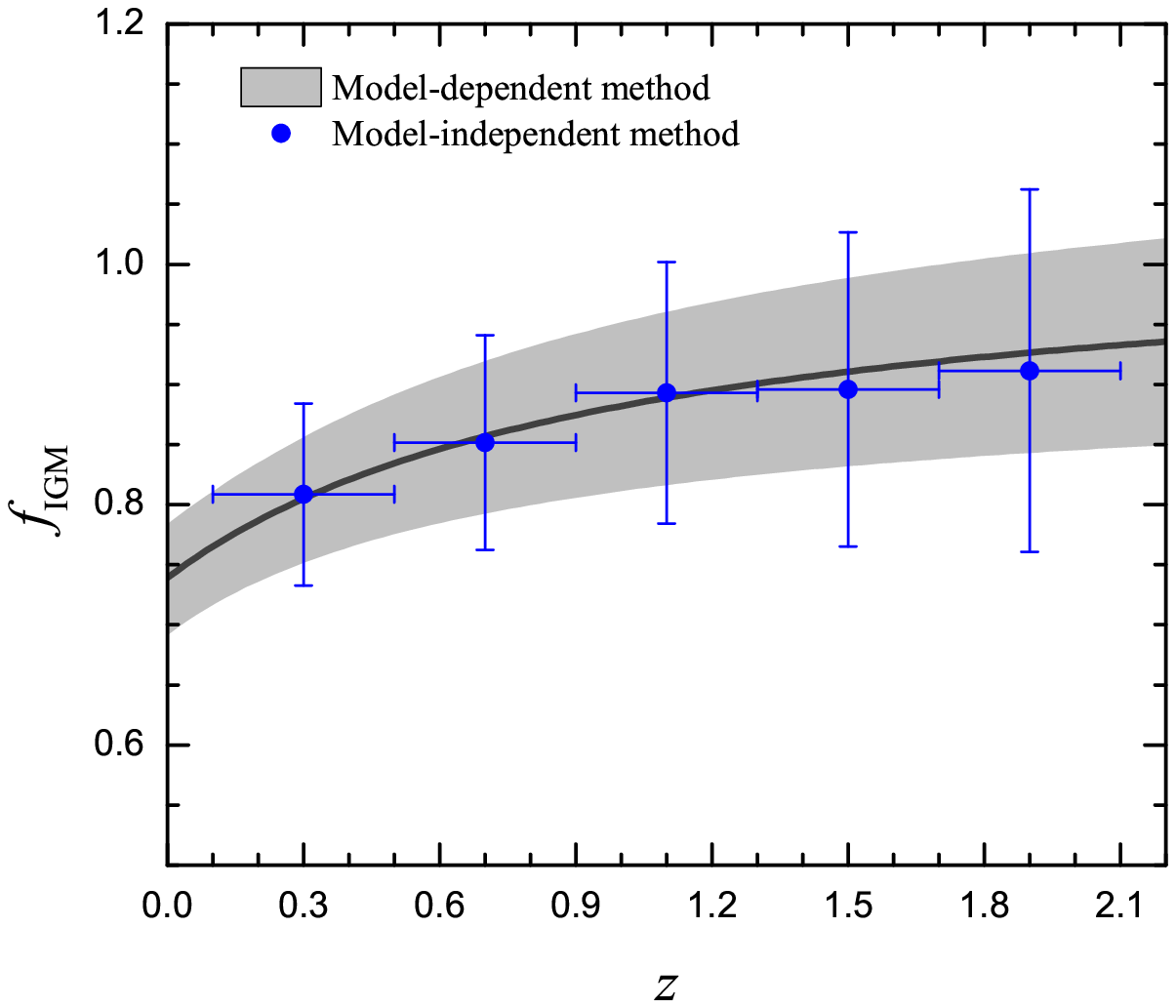}
%\vskip-0.2in
\caption{Left: 1D marginalized probability distributions and $1-3\sigma$ constraint contours of
$(f_{\rm IGM,0}, \alpha, \Omega_{m}, w_{0})$ in the $w$CDM model determined with the usual DM--$z$
method for 3000 simulated FRBs + 50 simulated $H(z)$ data. Right: $f_{\rm IGM}(z)$ function (solid line) with $1\sigma$
confidence region (shaded area) derived from 3000 simulated FRBs and 50 simulated $H(z)$ data with the usual DM--$z$ (model-dependent) method.
Blue dots correspond to the $f_{\rm IGM}(z)$ constraints at different redshifts derived from
our model-independent method for the same mock FRB and $H(z)$ data.}
\label{f6}
\end{figure}

\section{Summary and discussion}
\label{sec:summary}
The measured DM and $z$ of FRBs have been applied to study cosmology, but one happens to encounter the
strong degeneracy between cosmological parameters and the very uncertain baryon fraction in the IGM, $f_{\rm IGM}$.
In this paper, we propose that combining the FRB and
Hubble parameter $H(z)$ data in similar redshift ranges offers a new model-independent way to constrain the
evolution of $f_{\rm IGM}(z)$. Based on the ${\rm DM_{IGM}}(z)$ data derived from FRBs, we use the data-binning technique
to estimate the derivative of ${\rm DM_{IGM}}(z)$ with respect to $z$, ${\rm DM}'_{\rm IGM}(z)$. By combining
$H(z)$ and ${\rm DM}'_{\rm IGM}(z)$ at the same redshifts, we can directly determine $f_{\rm IGM}(z)$.
Through Monte Carlo simulations, we prove that the evolution of $f_{\rm IGM}(z)$ can be well inferred
from a sample of FRBs using our method.

The key issue in the idea of constraining $f_{\rm IGM}(z)$, however, is measuring the redshifts of a large
sample of FRBs. Essentially, the redshifts of FRBs can be measured by identifying their host galaxies or counterparts
in other electromagnetic wavelengths. With the help of very-long-baseline interferometry observations,
one may precisely localize their host galaxies, especially for dedicated observations on the repeating FRBs.
By promptly performing multi-wavelength follow-up observations after the FRB trigger, one may catch the associated
gamma-ray bursts \citep{2014ApJ...780L..21Z} or any other bright counterparts
\citep{2014ApJ...792L..21Y,2016ApJ...824L..18L,2016Natur.531..202S,2017ApJ...836L..32Z}.
It is encouraging that the upcoming CHIME and HIRAX radio arrays could detect $\sim10^{4}$ FRBs per year
\citep{2016SPIE.9906E..5XN,2018ApJ...863...48C}. In the next few years, a large sample of FRBs with redshift
measurements may become available, so the model-independent analysis of $f_{\rm IGM}(z)$ proposed here may be carried out.

As described above, given a cosmological model, one can directly obtain estimates of $f_{\rm IGM}(z)$
by using the usual DM--$z$ way. To investigate the importance of developing cosmology-free estimators,
we also run the usual DM--$z$ method to constrain $f_{\rm IGM}(z)$. We allow the parameters
$f_{\rm IGM,0}$ and $\alpha$ of the parameterized $f_{\rm IGM}(z)$ function to be free along with
the matter energy density $\Omega_{m}$ and the dark energy equation-of-state $w_{0}$ in the $w$CDM model,
and optimize these four free parameters by minimizing the $\chi^{2}$ statistic, i.e.,
\begin{equation}
\chi^{2}_{\rm tot}=\chi^{2}_{\rm DM}+\chi^{2}_{H}\;,
\end{equation}
where
\begin{equation}
\chi^{2}_{\rm DM}(\textbf{p}_{1}, \textbf{p}_{2})=
\sum_{i}\frac{\left[{\rm DM}^{\rm sim}_{\rm IGM}\left(z_{i}\right)-{\rm DM}^{w{\rm CDM}}_{\rm IGM}\left(z_{i};\; \textbf{p}_{1}, \textbf{p}_{2}\right)\right]^{2}}{\sigma^{2}_{{\rm tot},i}}
\end{equation}
and
\begin{equation}
\chi^{2}_{H}(\textbf{p}_{2})=
\sum_{i}\frac{\left[H^{\rm sim}\left(z_{i}\right)-H^{w{\rm CDM}}\left(z_{i};\; \textbf{p}_{2}\right)\right]^{2}}{\sigma^{2}_{H,i}}\;.
\end{equation}
Here $\textbf{p}_{1}=\{f_{\rm IGM,0}, \alpha\}$ and $\textbf{p}_{2}=\{\Omega_{m}, w_{0}\}$
stand for the parameters of the parameterized $f_{\rm IGM}(z)$ function and of the cosmological model, respectively.
${\rm DM}^{w{\rm CDM}}_{\rm IGM}(z)$ and $H^{w{\rm CDM}}(z)$ denote the theoretical values calculated from
the concerning parameters, ${\rm DM}^{\rm sim}_{\rm IGM}(z)$ and $H^{\rm sim}(z)$ are the simulated data
as explained earlier in Section~\ref{sec:simulation}, and $\sigma_{\rm tot}$ and $\sigma_{H}$ correspond to
the original errors of ${\rm DM}^{\rm sim}_{\rm IGM}(z)$ and $H^{\rm sim}(z)$, respectively.
Here the redshift distribution of FRBs is assumed to be the phenomenological model for GRB redshift distribution.
To ensure the final constraints are unbiased, we also repeat the simulation process 1000 times for each FRB + $H(z)$ data set
using different noise seeds. In the left panel of Figure~\ref{f6}, we display the confidence regions of
$(f_{\rm IGM,0}, \alpha, \Omega_{m}, w_{0})$ in the $w$CDM model determined with the usual DM--$z$ (model-dependent) method
for 3000 simulated FRBs + 50 simulated $H(z)$ data. The contours show that at the $1\sigma$ confidence level,
the best fits are ($f_{\rm IGM,0}=0.739^{+0.044}_{-0.047}$, $\alpha=0.286^{+0.106}_{-0.104}$,
$\Omega_{m}=0.325^{+0.041}_{-0.044}$, $w_{0}=-1.041^{+0.211}_{-0.307}$).
With the constraint results of $f_{\rm IGM,0}$ and $\alpha$, the derived $f_{\rm IGM}(z)$ function (solid line)
with $1\sigma$ confidence region (shaded area) is plotted in the right panel of Figure~\ref{f6}, where
we also present the $f_{\rm IGM}(z)$ constraints at different redshifts (blue dots) derived from our model-independent
method for the case of 3000 simulated FRBs and 50 simulated $H(z)$ data for the sake of comparison. One can see that
the derived constraints on $f_{\rm IGM}(z)$ from our model-independent method are somewhat weaker than
those of the usual DM--$z$ method. But it is worth pointing out that the $f_{\rm IGM}(z)$ constraints obtained from
the usual DM--$z$ method would become worse when considering the evolution of the dark energy equation-of-state.
What's more, the constraints on the evolving $f_{\rm IGM}(z)$ with our method are more robust and widely applicable,
as they do not depend on the cosmological model.

\acknowledgments
We are grateful to the anonymous referee for the useful suggestions which have helped us
to improve the presentation of the paper.
This work is partially supported by the National Natural Science Foundation of China
(grant Nos. 11603076, 11673068, 11725314, U1831122, 11505008, 11722324, 11603003, 11373014, and 11633001), the Youth Innovation Promotion
Association (2017366), the Key Research Program of Frontier Sciences (grant No. QYZDB-SSW-SYS005),
the Strategic Priority Research Program ``Multi-waveband gravitational wave universe''
(grant No. XDB23000000) of Chinese Academy of Sciences, the ``333 Project''
and the Natural Science Foundation (grant No. BK20161096) of Jiangsu Province,
and the Guangxi Key Laboratory for Relativistic Astrophysics.

%\bibliographystyle{JHEP}
%\bibliography{ms}

\providecommand{\href}[2]{#2}\begingroup\raggedright\endgroup

\end{document}